\input harvmac

\vskip 1cm

 \Title{ \vbox{\baselineskip12pt\hbox{}}}
 {\vbox{
\centerline{ Noncommutative Spheres and the AdS/CFT Correspondence  }
\centerline{   }  }}
\centerline{$\quad$ {Antal Jevicki, Mihail Mihailescu, Sanjaye Ramgoolam }}
\smallskip
\centerline{{\sl  }}
\centerline{{\sl Brown  University}}
\centerline{{\sl Providence, RI 02912 }}
\centerline{{\tt antal,mm,ramgosk@het.brown.edu}}
 \vskip .3in 
 
We present direct arguments for non-commutativity of
spheres in the AdS/CFT correspondence. The discussion is based
on results for the $S_N$ orbifold SCFT. Concentrating on three point 
correlations 
(at finite $N$) we exhibit a comparison with  correlations 
on a non-commutative sphere. In this manner an  essential signature
of  non-commutativity is  identified giving further support for the
original proposal of hep-th/9902059.
 
 
\Date{06/2000} 

\lref\freedmanmathur{D.Freedman, S.Mathur, A.Mathusis and L.Rastelli,
   Correlation functions in the \quad CFT(d)/AdS(d+1) correspondence, 
  Nucl.Phys.B546(1999): 96-118, hep-th/9804058 }
\lref\malda{ J. Maldacena, The large N limit of superconformal 
                             field theories and supergravity, 
                      Adv.Theor.Math.Phys.2: 231-252, 1998, 
                       hepth/9711200}
\lref\malstrom{ J. Maldacena, A. Strominger, AdS3 Black Holes and 
a Stringy Exclusion Principle,  JHEP 9812 005, 1998, hep-th/9804085}
\lref\jeram{A.Jevicki, S.Ramgoolam, Non commutative geometry from the
    ADS/CFT correspondence, JHEP 9904: 032, 1999, hep-th/9902059}
\lref\mijeram{A.Jevicki, M.Mihailescu, S.Ramgoolam, Gravity from CFT on
 $S^N(X)$ CFT: Symmetries and Interactions,Nucl.Phys.B577: 47-72, 2000, hep-th/9907144}
\lref\madore{J. Madore, The fuzzy sphere, Class.Quant.Grav. 9: 69-88, 1992}
\lref\gkp {S.S.Gubser,I.R.Klebanov and A.M.Polyakov, 
   Gauge Theory Corellators from Non-Critical String Theory, Phys.Lett.B428 :105-114, 1998, 
     hep-th/9802109 }
\lref\witten{ E. Witten, 
Anti-de-Sitter space and holography, Adv.Theor.Math.Phys.2:253-291, 1998, hep-th/9802150 } 	 
\lref\JevYo{A. Jevicki, T. Yoneya, Space-time uncertainty principle and conformal symmetry in D particle dynamics,
Nucl.Phys.B535:335-348,1998, hep-th/980506}
\lref\lunmat{O. Lunin, S. D. Mathur, Correlation functions for $M^N / S(N)$ orbifolds, hep-th/0006196 }
\lref\JevTo{A. Jevicki, A. van Tonder, Finite [Q oscillator] representation of 2-D string theory, 
Mod.Phys.Lett.A11:1397-1410, 1996, hep-th/9601058} 
\lref\HoRaTa{Pei-Ming Ho, Sanjaye Ramgoolam, Radu Tatar, Quantum space-times and finite N effects in 4-D 
SuperYang-Mills theories, Nucl.Phys.B573:364-376,2000, hep-th/9907145}
\lref\jlee{J. Lee, Three point functions and the effective lagrangian for the chiral primary fields in D = 4 supergravity
on $AdS(2) \times S^2$, hep-th/0005081} 
\lref\MSussT{ J. McGreevy, L. Susskind, N. Toumbas, Invasion of the giant gravitons from anti-de Sitter space,
 JHEP 0006:008,2000, hep-th/0003075} 
\lref\Yo{T.Yoneya , String Theory and Space-Time Uncertainty Principle,hep-th/0004074 }
\lref\myers{R.C.Myers,Dielectric Branes,hep-th/9910053 }
\lref\KT{D.Kabat and W.Taylor,Spherical membranes in Matrix theory, hep-th/
9711078 }
\lref\Holi{Pei-Ming Ho and M.Li,Fuzzy spheres in AdS/CFT and holography from 
noncommutativity, hep-th/0004072 }
\lref\BV{M.Berkooz and H.Verlinde,
Matrix Theory,ADS/CFT and the Higgs-Coulomb Equivalence ,
hep-th/9907100}
\lref\BJL{D.Berenstein,V.Jejjala and R.G.Leigh,hep-th/0005087}

\lref\cz{ T. L. Curtright and C.  Zachos,
 Deforming maps for quantum algebras, Phys. Lett B243, 237-244, 1990 } 
\lref\Pod{ P. Podles, Quantum spheres,
 Lett. Math. Phys. 14, 193-202 (1987) } 
\lref\mo{ T. Masuda, K. Mimachi, Y. Nakagami, M. Noumi, K. Ueno, 
 Representations of quantum groups and a
 q-analogue of orthogonal polynomials, C.R. Acad. Sci. Paris 307, 
 559-564 } 
\lref\stein{H. Steinacker, Quantum Anti-de Sitter space and sphere at
roots of unity, 1999, hep-th/9910037}
\lref\ferr{S. Ferrara, M.A. Lledo, Some aspects of deformations of
supersymmetric field theories, JHEP 0005:008, 2000, hep-th/0002084} 
\lref\suzu{ H. Suzuki, Rational scalar field theories on quantum spheres,
Prog.Theor.Phys.91:379-391, 1994} 
\lref\schi{R. Schiappa, Matrix strings in weakly curved background fields, 2000, hep-th/0005145} 
\lref\klim{H. Grosse, C. Klimcik, P. Presnajder, Towards finite quantum field theory in noncommutative geometry, 
Int.J.Theor.Phys. 35: 231-244, 1996, hep-th/9505175 }
\lref\grat{J. Gratus, An introduction to the noncommutative sphere and some extensions, 1997, q-alg/9710014}

\def\lam{\lambda}

\newsec{Introduction}
The AdS/CFT correspondence $\malda,\gkp,\witten$  provides a constructive
approach to supergravity, and closed string theory in curved backgrounds. 
Its basis is the large $N$ expansion of Yang-Mills type theory which turns 
into a loop expansion of gravity. The existence of a deductive procedure 
has prompted questions concerning any modification that the emerging gravity 
could exhibit. In particular a proposal was made in $\jeram$, $\mijeram$, that 
the curved SUGRA background $AdS \times S$ is non-commutative with the 
noncommutativity parameter given by ${1\over N}$. This follows an earlier 
proposal for q-deformation given in a framework of a simple matrix model $\JevTo$. 
The extension to higher spaces was further discussed in $\HoRaTa$.
\par
The non-commutativity of $AdS \times S$ space naturally incorporates the 
exclusion principle of $\malstrom$ and stands to have important implications 
on the physics of black holes. 
It also conforms to a general principle of $\Yo$ that string/M-theory inherently
contains a space-time uncertainty( see $\JevYo$ ). Recently, a physical 
argument for 
the noncommutativity and the associated cutoff was given 
in $\MSussT$ based on a study 
of brane motions on spheres. This discussion involves a 
mechanism $\myers$, $\KT$ by which 
gravitons are polarized into extended spherical membranes which lead to non-commutativity 
$\Holi$. Other  examples are given in $\BV$, $\schi$, $\BJL$, $\stein$, $\ferr$.
\par
 It is clearly important to further clarify the nature of non-commutativity 
in $AdS \times S $ spaces. In comparison with the noncommutativity  induced  by
an external B-field for  the origin of noncommutativity in $AdS \times S$ is 
less trivial to exhibit. Since it  involves the closed string
coupling ${1\over N}$,  it represents a nonperturbative
phenomenon.
\par
In this paper, we describe further evidence for the above noncommutativity
in AdS/CFT. The discussion is based on the $S_N$ orbifold model that already
served as the basis for arguments presented in $\jeram$, $\mijeram$. In this 
model, 
one is able to perform explicit calculations of three point interactions 
at finite 
$N$ and study their behavior. In this way, we exhibit an explicit 
signature for non-commutativity of the corresponding spheres.
\par
The content of the paper is as follows. In section 2, we summarize the
finite $N$ results for three-point correlations in the orbifold CFT 
and discuss their properties.
We then
review the (super)gravity calculations in commutative space-time in section 3
and proceed to evaluate the modifications due to a non-commutative sphere in
section 4.We exhibit certain agreement with the SCFT study.     

\newsec{Results from $S_N$ orbifold }
\par
In this section, we review the results of CFT dual to the gravity in the case of 
$AdS_3 \times S^3$ obtained in $\mijeram$. We also use the extension of 
these results to the nonextremal case that can be read off from recent work of
$\lunmat$. 
\par
The SCFT in question is defined on symmetric  product  
$S^N(M)$, where $M$ is either $T^{4}$ or $K3$ . 
The field content of the theory consists of: $4N$ 
real free bosons $X^{a \dot{a}}_{I}$ representing the 
coordinates of the torus for example and their superpartners 
$4N$ the fermions $\Psi^{\alpha\dot{a}}_{I}$, where $I=1,..,N$, 
$\alpha,\dot{\alpha} =\pm$ are the spinorial $S^{3}$ 
indices, and $a,\dot{a}=1,2$ are the spinorial indices on $T^{4}$.
In essence, the field content of the theory is determined to be $4 N$ real 
free bosons and  $2 N$ Dirac free fermions, giving a central charge $c=6 N$. 
One has left and right superconformal symmetry with the corresponding currents. 
The lowest modes of this currents ,namely 
$ \{ L_{0,\pm 1},G_{\pm  {1 \over 2}}^{\alpha a},J_{0}^{\alpha\beta} \}$
together with their right counterparts generate the 
$SU(2|1,1)_{L} \times SU(2|1,1)_{R}$ symmetry.These according to the $AdS/CFT$ 
correspondence translate into the superisometries of the $AdS_{3}\times S^{3}$
space-time. 
In addition, one has other symmetries commuting 
with the previous set for example the ones related to global $T^{4}$ 
rotations. Even 
though the underlying CFT on $T^4$ is very simple, the complexity is 
given by the  non-trivial implementation of the $S_N$ symmetry of the 
orbifold.This symmetry is an analogue of U(N) gauge symmetry in this
context .Physical observables analogous to traces are now given by
$S_N$ invariants . In particular the complete set of chiral primary 
operators was given in $\jeram $, $\mijeram$. A fundamental role is played by 
the twist operators that impose the twisted the boundary conditions : 
\eqn\bcond {
X_{I}(z\,e^{2 \pi i}, \bar{z}\,e^{-2 \pi i})=X_{I+1}(z,\bar{z}), \; 
I=1..n-1, \;
X_{n}(z\,e^{2 \pi i}, \bar{z}\,e^{-2 \pi i})=X_{1}(z,\bar{z})
}
for the n-twisted sector of the theory.First a construction of $Z_n$ twist 
operators is given in terms of which the 
$S_N$ invariant chiral primary operators are constructed
after appropriately averaging over $S_N$.
 
In the correspondence with gravity on $ADS_3 \times S^3$ one 
achieves a one-one correspondence with single particle states. 
 
\par
The computation of two and three point functions (for extremal momenta) was 
given in $\mijeram$ and has recently been extended in $\lunmat$.   
We present the three-point function for chiral primaries $O_{n}^{(0,0)}$, where
$n$  denotes the length of the cycle (twist) used to construct the operator.
 The 
index $n$ is also identified with $l$ of the angular momentum on $S^3$ in 
gravity as $n=l+1$
(let us recall that the isometry group for $S^3$ is $SO(4)=SU(2)\times SU(2)$ and that 
$l$ is an angular momentum in the diagonal $SU(2)$). The correlation
functions in terms of $n=l_1+1$, $k=l_2+1$ and 
$n+k-1=l_3+1$ for the case $l_3=l_1+l_2$ (extremal) are:
\eqn\chiraop{
\langle\,O_{n+k-1}^{(0,0)\,\dagger}(\infty)\,
O_{k}^{(0,0)}(1)\,O_{n}^{(0,0)}(0)\rangle =  
\left({{(N-n)!\,(N-k)!\, (n+k-1)^3 } \over 
{(N-(n+k-1))!\,N!\,n\,k}}\right)^{{1 \over 2}} 
}
This expressions for the three point correlation functions obtained in $\mijeram$
 contains two types of factors: one that is solely dependent on the angular 
 momenta with no  $N$ dependence and the other with explicit $N$ dependence .
 It is the latter type  that we concentrate on as it contains direct implications on the
 non-commutative nature of the spacetime.
 \par
  The above result stands for the extremal $l_1+l_2=l_3$  case. We would now like to 
 extend its main features to the general non-extremal case.
 We are guided
 by the following two facts. First are the results  of $\lunmat$ 
 which were done for a simpler  bosonic orbifold but are found for general $l$`s. 
 Furthermore, as can be seen from the original discussion
 of$\mijeram$ the important  $N$ dependence relevant to the cutoffs
 essentially originates from combinatorial properties of 
permutation group. Correspondingly, for general  angular momenta 
$l_{1,2,3}$, we can expect the factor coming from $S_N$ permutations
of the form:
\eqn\termnon{
{\left((N-1-l_1)! (N-1-l_2)! (N-1-l_3)!\right)^{1\over2} \over (N-1-{{l_1+l_2+l_3}\over2})!}
}
It is this factor that contains the signature of non-commutativity.
 Firstly,
one sees  the "exclusion principle" that, for a non-zero 
correlation function, any individual  angular momenta $l$ should not
exceed  the bound N. There is  a further bound on the sum
of angular momenta characteristic of fusion rules of WZW
 models or of $SU_q(2)$ at roots of unity. But as we  will see below,
 the full factorial 
form will appear to be present in the correlations to be evaluated 
on a non-commutative sphere.

\par

\newsec{Field theory on $AdS \times S$} 
\par
We will now perform two parallel calculations,one with the 
standard (commutative)
sphere and the other with the non-commutative (fuzzy) sphere. 
The purpose is to 
directly compare the results and see that the later case exhibits the factorial
terms that were featured in the orbifold calculation.
\par
In evaluating the  correlation functions in SUGRA  one has the
following two-step process. The $AdS$ dependence is projected to the boundary
of the $AdS$ space-time using the bulk to boundary propagator. For the sphere 
one expands in terms of spherical harmonics.
The signature of noncommutative space that we are going to exhibit is 
associated with the $S$ (sphere) part of this calculation. The $AdS$ part 
does not lead to such a characteristic behavior and in much of what will be 
presented can be ignored. Indeed we should expect the essential features of the cutoff
 seen in the boundary correlators to be associated with the sphere part, since the models 
 of q-deformed $ADS$ considered in \jeram\Holi\
have the property that the boundary is 
 commutative. Other aspects like space-time uncertainty are manifested by the 
 non-commutative ADS part. 
We will in the present section
summarize the calculation of correlation functions
  in the commuting case and then in the next section 
repeat the analogous calculations in the non-commutative case. 
\par
Consider for simplicity a massless  scalar field with cubic interaction:
\eqn\lagr{
S= \int d x \sqrt{g}((\partial \Phi)^2+ \lambda \Phi^3 + \dots) 
  }
where the integral is over the  $AdS \times S $ space, $g$ is the 
corresponding metric and $\lam$ represents the coupling constant for 
the cubic interaction. 
In studying field theory on products $AdS \times S$ space, the wavefunctions 
factorize into $AdS$ and $S$ components
$$
\Phi = \sum\Phi (\rho , t , \phi ) \Psi_I (\vec{n} )
$$
where $ \Psi_I (\vec{n} )$ denote the spherical functions on the sphere $S$.  For the case $S_3$, one has the well known $D_{mm'}^{\ell} (\theta , \varphi , \psi )$ functions.  For three point functions, one has schematically
$$
\langle \Phi \Phi \Phi \rangle_{AdS} \, \langle\Psi_{\ell_{1}} \Psi_{\ell_{2}}
\Psi_{\ell_{3}}\rangle_S
$$
The sphere contributions (on which we will concentrate) 
exhibit typically the Clebsch-Gordon coefficients
$$
\pmatrix{\ell_1 & \ell_2 & \ell_3\cr
m_1&  m_2&  m_3 \cr}
\pmatrix{\ell_1 & \ell_2 & \ell_3\cr
m_1' & m_2' & m_3' \cr}
F \left( \ell_1 , \ell_2 , \ell_3 ; N\right)
$$
and a factor $F(\ell ; N)$ solely dependent on the magnitudes of angular momenta and 
$N$.  In the non-commutative case, this is the structure of a star product
$$
Y^{\ell_{1}} * Y^{\ell_{2}} = \sum \left( Clebsch - Gordon \right) f \left( \ell_1 \ell_2 \ell_3 ;  N\right) Y^{\ell_{3}}
$$
For studying the form of the ``fusion coefficient", $F(\ell_1 \ell_2 , \ell_3 :N)$ , it is sufficient to consider the reduction of the sphere $S_3$ to $S_2$.  One knows that in particular
$$
D_{m m' = 0}^{\ell} = Y_{\ell m} (\theta , \phi)$$
i.e. we have  spherical harmonics on $S_2$.  The nature of non-commutativity is essentially the same for the $AdS_3 \times S_3$ space or $AdS_2 \times S_2$ which is its $U(1) \times U (1)$ coset.  
So in what follows for simplicity of the calculation,  we will discuss the latter.
\par 
We use the following representation for spherical harmonics:
\eqn\harm{
Y^I = \Omega^I_{i_1 \dots i_l} {{x^i_1 \dots x^{i_l}} \over \rho^l},
}
where $\Omega^I$ is a traceless, symmetric, $l$-index tensor with indices 
$i_k=1 \dots 3$, $x^i$ are the coordinates of the three dimensional flat 
space and $\rho=\sqrt{(x^1)^2+(x^2)^2+(x^3)^2}$. $Y^I$ is an eigenvector of
the sphere laplacian with eigenvalue $l(l+1)$. We also assume that the tensors 
$\Omega^I$ 
are normalized in the sense: 
$\Omega^I_{i_1 \dots i_l} \Omega^J_{i_1 \dots i_l}=\delta^{I J}$ for equal 
index number and $0$ otherwise. By straightforward computations presented in the appendix 1 we 
obtain the following expressions for the product of two harmonics 
((I,$l_1$), (J,$l_2$) indices) integrated over the sphere:
\eqn\prodtw{
\langle Y^I Y^J \rangle\equiv{1 \over {4 \pi}}\int_{S^2} Y^I Y^J = {{\pi^{1\over2} \Gamma(l+1)} \over 
{2^{l+1} \Gamma(l+{3\over2})}}\delta^{I J}, 
 }
For the product of three harmonics ((I,$l_1$), (J,$l_2$), (K,$l_3$) indices)
we obtain:
\eqn\prodthr{
\langle Y^I Y^J Y^K\rangle\equiv{1 \over {4 \pi}}\int_{S^2} Y^I Y^J Y^K={{\pi^{1\over2} 
\Gamma(l_1+1) \Gamma(l_2+1) \Gamma(l_3+1)} \over 
{2^{{\Sigma \over 2}+1} \Gamma(\alpha_1+1)\Gamma(\alpha_2+1)
\Gamma(\alpha_3+1)  \Gamma({\Sigma+3 \over 2})}} C^{I J K}
}
where $\Sigma=l_1+l_2+l_3$, $\alpha_1={1 \over 2}(-l_1+l_2+l_3)$, 
$\alpha_2={1 \over 2}(l_1-l_2+l_3)$, $\alpha_3={1 \over 2}(l_1+l_2-l_3)$ and:
\eqn\clebsch{
C^{I J K}=\Omega^I_{i_1\dots i_{\alpha_3}k_1\dots k_{\alpha_2}}
\Omega^J_{i_1\dots i_{\alpha_3}j_1\dots j_{\alpha_1}}
\Omega^J_{j_1\dots j_{\alpha_1}k_1\dots k_{\alpha_2}}
}
We multiply the spherical harmonics with appropriate factors in order to 
normalize them. After we integrate over the sphere we obtain the following 
$AdS_2$ action:
\eqn\redact{
S=\int_{AdS_2} d^2 x \sqrt{g}(\partial \phi_I \partial \phi_I+l(l+1)\phi_I \phi_I + 
\lam A^{I J K}\phi_I \phi_J \phi_K)
}
where $g$ is now the $AdS_2$ metric and:
\eqn\threo{
A^{I J K}={{\pi^{-{1 \over 4}} 2^{1 \over 2}\left(\Gamma(l_1+1)\Gamma(l_1+1)
\Gamma(l_3+1) \Gamma(l_1+{3 \over 2})\Gamma(l_2+{3 \over 2})
\Gamma(l_3+{3 \over 2})\right)^{1 \over 2}}\over{\Gamma({\Sigma+3 \over 2})
\Gamma(\alpha_1+1)\Gamma(\alpha_2+1)
\Gamma(\alpha_3+1)}}.
}
The action in $AdS_2$ can be related to the boundary action using the 
procedure and expressions developed in $\witten$, $\freedmanmathur$ 
(see also $\jlee$). The two- and three- point expressions for the constants 
in the correlation functions of the boundary operators 
${\it O}_I$ (corresponding to $\phi_I$) are:
\eqn\correl{\eqalign{
& \langle {\it O}_I {\it O}_J \rangle=\delta_{I J} 
{\Gamma(\Delta_I+1) \over {\pi^{1 \over 2} \Gamma(\Delta_I-{1\over2})}}, \cr
& \langle {\it O}_I {\it O}_J {\it O}_K \rangle= -{\lam A^{I J K} \over 2\pi}
{{\Gamma({{-\Delta_I+\Delta_J+\Delta_K}\over 2})
 \Gamma({{\Delta_I-\Delta_J+\Delta_K}\over 2})
 \Gamma({{\Delta_I+\Delta_J-\Delta_K}\over 2})
 \Gamma({{\Delta_I+\Delta_J+\Delta_K-2}\over 2})} \over
{\Gamma(\Delta_I-{1 \over 2})\Gamma(\Delta_J-{1 \over 2})
 \Gamma(\Delta_K-{1 \over 2})}},
}}
where $\Delta$ for each operator is $l+1$. We redefine the operators such 
that the constant in the two-point function is $\delta_{I J}$. After introducing all the factors, those coming 
from normalization and $A^{I J K}$, we obtain the following expression for the
three-point correlation functions:
\eqn\scalar{\eqalign
{ & \langle {\it O}_I {\it O}_J {\it O}_K \rangle=-{\lam C^{I J K} \over 
(2\pi)^{1\over2}}\left({(l_1+{1\over2})(l_2+{1\over2})(l_3+{1\over2})}\over
 {(l_1+1)(l_2+1)(l_3+1)}\right)^{1\over2}{{\Gamma(\alpha_1+{1\over2})
\Gamma(\alpha_2+{1\over2})\Gamma(\alpha_3+{1\over2})}
\over {\Gamma(\alpha_1+1)\Gamma(\alpha_2+1)\Gamma(\alpha_3+1)}} \times \cr
& \qquad \qquad \times {\Gamma({{\Sigma-1}\over2}) \over \Gamma({{\Sigma+3}\over2})}.\cr
}}
We observe that in our case the scalar has a mass given by
the sphere laplacian $\sqrt{l(l+1)}$. In the case of $AdS_2 \times S_2$ gravity, much of the 
analysis is in terms of chiral primary fields. The chiral primary fields are 
combinations of fields coming from four dimensional gravity that have the lowest possible 
$AdS_2$ mass for a given $l$: $\sqrt{l(l-1)}$. We can assume
that we do not start with such a simple theory as in $\lagr$, but with one 
that after sphere reduction leads to the lowest mass, appropriate for a chiral primary 
field $\witten$. For such a field, the corresponding $\Delta$ is $l$.  
In such a theory, we also consider a simple cubic interaction and 
obtain a qualitative picture for the sphere reduction. In this case, the three-point 
correlation functions for chiral primaries operators obtained in the end are simpler:
\eqn\chiras{
\langle {\it O}_I {\it O}_J {\it O}_K \rangle=-{\lam C^{I J K} \over 
(2\pi)^{1\over2}}{{4\left((l_1^2-{1\over4})(l_2^2-{1\over4})
(l_3^2-{1\over4})\right)^{1\over2}}\over{\alpha_1\alpha_2\alpha_3(\Sigma^2-1)}}.
}  
We observe that the correlation functions in the case of chiral primary 
$\chiras$ is much simpler than $\scalar$. This is one of the features observed 
in all gravity cases of $AdS_p \times S^p$ reduction for chiral primary operators. 
The cancellation of factors appears between those coming from sphere and those
coming from $AdS$ spaces. We also note that $\chiras$ gives a divergent result
for extremal cases like $l_1+l_2=l_3$. The divergences are due to our 
simplified model and they disappear in a realistic model. In the case of gravity, 
both the quadratic and cubic terms contain higher derivatives of the fields and 
these are responsible for both lower mass and consistent three-point correlation 
functions. 
\newsec{ Field theory on non-commutative $AdS_2 \times S^2$ }
\par
 We will now repeat the calculation of the above section, replacing 
 the  sphere by its noncommutative counterpart.
 In our previous work we have argued for non-commutative 
(q-deformation of)   
space-time in the  context of the AdS/CFT correspondence.
 The q-deformed two-sphere can be defined as a quotient
 of $SU(2)_q$ \mo\ and belongs to the classification 
 of \Pod. 
 There are also   transformations  between 
 q-spheres with manifest $SU(2)_q$ symmetry
 and spheres with manifest $SU(2)$ symmetry. 
 For generic $q$ this takes the form of 
 a connection between the classical sphere and 
 the q-sphere. For roots of unity this takes the form 
 of a connection between the q-sphere and the fuzzy sphere. 
 The technical reason for these connections is essentially 
 the deformation maps between $U_q$ generators and the generators
 of the classical symmetry discussed in \cz. 
 Applying these transformations to both the algebra of functions 
 on the  sphere and to the symmetry generators
 acting on the algebra gives a transformation between 
  $q$-sphere with $U_q$ symmetry
 and classical sphere with $U(SU(2))$ symmetry for 
 generic $q$. This can be expected to 
  lead, at roots of unity, to a transformation 
 between fuzzy sphere and $q$-sphere. Indeed  $q$-spheres
 at roots of unity are know to admit finite  $U_q$ covariant 
truncations \suzu.
 The transformation is a non-commutative version of a diffeomorphism
 which should be a symmetry in these applications of 
  quantum spheres to non-commutative gravity. In the following we 
 work with the fuzzy sphere.  

 We first review the definition and  properties of the fuzzy sphere
giving formulae  for the integration in parallel with the 
commutative case. This will lead to calculation of all 
the relevant quantities such as the normalization constants 
$\prodtw$ and the three harmonic interaction $\prodthr$. 
\par
The definition of the fuzzy sphere (for reviews see $\madore$, 
$\klim$, $\grat$) is given 
in terms of an algebra of polynomials in $X^i$, 
$i=1\dots3$ subject to the following constraints:
\eqn\constr{\eqalign
{
& [X^i,X^j]=i\epsilon_{i j k} X^k , \cr
& (X^1)^2+\dots+(X^3)^2=\rho^2, \cr
}}
where $\rho^2$ is a constant equal to ${{N^2-1}\over4}$ 
($N$ is a positive integer measuring the
fuzziness of the sphere). 
Such a deformation preserves the $SO(3)$ symmetry of the sphere. 
We can represent the $X$'s (and the algebra) as 
hermitian operators in the $SO(3)$ representation having spin 
${{N-1}\over2}$. As such, the coordinates are now hermitian 
$N\times N$ matrices. In this representation we define  
the integral over the sphere as:
\eqn\inttr{
{1 \over 4\pi} \int_{S^2}(\dots) \rightarrow {1\over N}Tr(\dots)
}
where $Tr$ is the trace in the ${{N-1}\over2}$ representation and $\dots$ mean a function on the sphere. It is also 
straightforward to represent the $su(2)$ symmetry (generators $J_i$) in this algebra:
\eqn\sualg{
J_i f(X)= [X^i,f(X)],\qquad i=1\dots3
}
The spherical harmonics are constructed in the same way as in commutative sphere
$\harm$ replacing the commutative coordinates $x^i$ with the noncommutative ones
$X^i$. It is straightforward to prove that the vector space of symmetric traceless polynomial of degree $l$ is left invariant by the $su(2)$ generators
and that it forms an irreducible representation with highest weight $l$. 
\par
The symmetric polynomials in $X$ of any degree appear in the Taylor 
series of $exp(i J X)=exp(i J_i X^i)$, where $J$'s are regular commutative numbers.
In order to compute the normalization constants and the three point interaction
we construct the following quantity:
\eqn\tracess{
I( J^1, J^2, J^3 )={1\over N} Tr ( e^{i J^1 X} e^{i J^2 X} e^{i J^3 X} )
}    
We can extract from this the trace of the product of two and three symmetric 
polynomials as:
\eqn\trsse{\eqalign
{
& {1\over N}Tr(X^{(i_1} \dots X^{i_{l_1})} X^{(j_1} \dots X^{j_{l_2}) })=
\partial_{J^{1}_{i_1}}\dots \partial_{J^{1}_{i_{l_1}}}
\partial_{J^{2}_{j_1}}\dots \partial_{J^{2}_{j_{l_2}}} I(J^1, J^2, 0)|_{J^{1, 2}=0}, \cr
& {1\over N}Tr(X^{(i_1} \dots X^{i_{l_1})} X^{(j_1} \dots X^{j_{l_2})} 
X^{(k_1} \dots X^{k_{l_3})})= \cr
& \quad \partial_{J^{1}_{i_1}}\dots \partial_{J^{1}_{i_{l_1}}}
\partial_{J^{2}_{j_1}}\dots \partial_{J^{2}_{j_{l_2}}}
\partial_{J^{3}_{k_1}}\dots \partial_{J^{3}_{k_{l_3}}}I(J^1, J^2, J^3)|_{J^{1 2 3}=0},
\cr
}}  
where $(\dots)$ means the symmetrized product of $X$'s.
\par
The evaluation of $I( J^1, J^2, J^3 )$ can be done if we note that the RHS of
$\tracess$ is the trace of a product of three $SO(3)$ rotations with parameters
$J^{1,2,3}$. The product of three rotations is itself a rotation with a parameter
$J=J(J_1, J_2, J_3)$:
\eqn\rot{
e^{i J X}=e^{i J^1 X} e^{i J^2 X} e^{i J^3 X}, 
}   
and the trace of this operator can be evaluated easily in a basis where $JX$ is 
diagonal as:
\eqn\ttrt
{
I(J) \equiv I( J^1, J^2, J^3 )={1\over N} {sin({{J N}\over 2}) \over sin({J\over2})} 
}    
The dependence of $J$ (or rather $cos({J\over2})$) on $J^{1,2,3}$ can be easy 
computed (see appendix 2) and we list here the result:
\eqn\jcj{\eqalign
{
& cos({J\over2})=cos({J^1\over2}) cos({J^2\over2}) cos({J^3\over2})-
cos({J^1\over2}){sin({J^2\over2})\over J^2}{sin({J^3\over2})\over J^3}-
cos({J^2\over2}){sin({J^3\over2})\over J^3}{sin({J^1\over2})\over J^1}\cr
& \qquad -cos({J^3\over2}){sin({J^1\over2})\over J^1}{sin({J^2\over2})\over J^2}+
{sin({J^1\over2})\over J^1}{sin({J^2\over2})\over J^2}{sin({J^3\over2})\over J^3}
(J^1 \times J^2)J^3 \cr
}}
The cubic interaction in the case of fuzzy sphere introduces an additional subtlety,
namely:
\eqn\subt{
{1\over N}Tr(Y^I Y^J Y^K) \neq {1\over N}Tr(Y^I Y^K Y^J)
}
Because of this, some of the properties of cubic interaction we 
find in commutative case, are not there in the noncommutative case. In 
particular, we loose the appearance 
of the Clebsch-Gordon coefficients in the cubic interaction. This asymmetry is 
not present in the case $\phi^3$ interaction and not even in the case of 
$\phi_1^2 \phi_2$, but it is present in the case $\phi_1 \phi_2 \phi_3$ type 
interaction, where $\phi_{1,2,3}$ are three different fields. 
We like to preserve those properties of interaction, as they seem to be present in 
the CFT $\mijeram$, and we change the definition of the integration over the fuzzy
sphere by replacing the trace over the product of harmonics with the trace over 
the symmetric product of harmonics. The change amounts in the end in dropping 
the last factor appearing in the expression of $cos({J\over2})$ $\jcj$, the only 
one not symmetric in $J^{1,2,3}$. After this change, we remain with 
$J$ depending on the following variables only: $|J^1|$, $|J^2|$, $|J^3|$, 
$J^1 J^2$, $J^2 J^3$ and $J^1 J^3$, where $|J|=\sqrt{J^2}$. 
\par
The method used for the evaluation of the two- and three- interaction for harmonics
is given in $\trsse$. Spherical harmonics come with polynomial in $X$'s that are both
symmetric and traceless. The traceless property of polynomials is shifted to the 
traceless of partial derivatives in $J^1$, $J^2$ and $J^3$ $\trsse$ and as such, leads 
after setting  $J$'s to $0$ to the following expression (we denote by
$\tilde{A}^{(i) (j)}$ and $\tilde{A}^{(i) (j) (k)}$ the LHS of the equations 
$\trsse$ with the property that the indices $i$'s, $j$'s and $k$'s are also 
traceless):
\eqn\trsee{\eqalign
{
& \tilde{A}^{(i) (j)}=\delta_{l_1 l_2}{1\over 2^{2 l_1}}
({\partial\over \partial(cos({J\over2}))})^{l_1} I(J)|_{J=0} 
(\delta^{i_1 j_1}\dots\delta^{i_{l_1} j_{l_1}}+\dots), \cr
& \tilde{A}^{(i) (j) (k)}={1 \over 2^{l_1+l_2+l_3}}
({\partial\over \partial(cos({J\over2}))})^{{l_1+l_2+l_3}\over2} I(J)|_{J=0}
(\delta^{i_1 j_1}\dots\delta^{i_{\alpha_3+1} k_1}\dots+\dots) \cr
}}
The $\dots$ in the last equations mean all possible contractions between
indices $i$'s and $j$'s, $j$'s and $k$'s and $i$'s and $k$'s. We have also to 
specify:
\eqn\spre{
({\partial\over \partial(cos({J\over2}))})^l I(J)|_{J=0}={(N+l)!\over
{N (N-1-l)!(2l+1)!!}},
}
and we give a derivation for this in the appendix. The final results 
for harmonics can be written in terms of those obtained in the commutative 
case (we denote the corresponding terms for the fuzzy sphere as 
$\langle Y^I Y^J \rangle_N$ and $\langle Y^I Y^J Y^K\rangle_N$):
\eqn\prodNtw{\eqalign
{& \langle Y^I Y^J \rangle_N={(N+l_1)!\over {(2\rho)^{2l_1} N (N-l_1-1)!}}
 \langle Y^I Y^J \rangle, \cr
& \langle Y^I Y^J Y^K \rangle_N={(N+{{l_1+l_2+l_3}\over2})!\over 
 {(2\rho)^{l_1+l_2+l_3} N (N-1-{{l_1+l_2+l_3}\over2})!}}
 \langle Y^I Y^J Y^K\rangle, \cr
}}
where  $\rho$ is given $\constr$. In the $N \rightarrow \infty$ limit, 
the factors coming from the fuzzy sphere go to $1$ and the results for the commutative sphere are obtained.

\par
We have now prepared all the necessary ingredients to proceed to our 
main calculations and evaluate correlation functions  associated with
gravity on $AdS_2 \times S^2_{fuzzy}$. 
\par
Consider now the following action:
\eqn\redact{
S=\int_{AdS_2} d^2 x \sqrt{g} {1\over N}Tr(\partial \phi \partial \phi+
J_i \phi J_i \phi+\lam \phi^3)
} 
 where the integration over a (commutative) sphere is 
replaced by the $Tr$ over symmetric product of functions defined on the
non-commutative (fuzzy) sphere. Expanding in terms of associated spherical
harmonics, we obtain the 
same action on $AdS_2$ as in $\redact$ but with $A^{I J K}$ replaced by:
\eqn\aaa{
A_N^{I J K}={{\left((N-1-l_1)! (N-1-l_2)! (N-1-l_3)!\right)^{1\over2}}\over 
(N-1-{{l_1+l_2+l_3}\over2})!} 
{{ N^{1\over2} (N+{{l_1+l_2+l_3}\over2})!}\over
{\left((N+l_1)! (N+l_2)! (N+l_3)!\right)^{1\over2}}} A^{I J K}
}
After we reduce the theory on $AdS_2$ in the same way as before we obtain 
the result. The correlation functions in the case of
non-commutative sphere are equal with those in the case of commutative 
sphere multiplied
with the same factor as in $\aaa$. We give below the result for the chiral primaries 
correlation functions in this case (see $\chiras$):
\eqn\cirasn{\eqalign 
{
&
\langle {\it O}_I {\it O}_J {\it O}_K \rangle_N=-{\lam C^{I J K}\over 
(2\pi)^{1\over2}}
{{\left((N-1-l_1)! (N-1-l_2)! (N-1-l_3)!\right)^{1\over2}}\over (N-1-{{l_1+l_2+l_3}\over2})!} 
\times \cr
& \quad \times
{{N^{1\over2} (N+{{l_1+l_2+l_3}\over2})!}\over
{\left((N+l_1)! (N+l_2)! (N+l_3)!\right)^{1\over2}}}
{{4\left((l_1^2-{1\over4})(l_2^2-{1\over4})
(l_3^2-{1\over4})\right)^{1\over2}}\over{\alpha_1\alpha_2\alpha_3(\Sigma^2-1)}}.\cr
}}
\par
This expression contains the essential features relevant for comparison with
the CFT. One has the $SU(2)$ symmetry, that was represented as R-symmetry 
in CFT. The result exhibits a cutoff at $l_{1,2,3}=N$ and also at 
${{l_1+l_2+l_3}\over2}=N$. Most significantly the overall factor 
that we see in the noncommutative (fuzzy)
sphere result is
of identical form to the corresponding factor given by the $S_N$ orbifold . 
This is clear evidence for  non-commutativity in AdS/CFT. The 
characteristic features of the factorial terms describing this
non-commutativity are seen to be captured by the sphere. It is relevant to 
stress
that the $AdS$ contribution is not of the same form, which is understandable 
as we argued at the beginning of section 3 in terms of the commutative nature
 of the boundary in deformed ADS.  It should be stated
that  the  above expressions are not exactly equal in form to those 
of the finite N CFT. First the computation of this section is done for a 
simplified model.
In addition, 
the gravitational coupling  being given by $1/N$  means that  loop effects will
also contribute to the final $N$ dependence. Finally deformations of AdS will
also imply corresponding $N$ dependence. But we expect that these two  effects,
while providing  $N$ dependence, do not lead to a contribution 
of the form  that we have identified and associated with
a fuzzy sphere.
 
\bigskip

\noindent{\bf Acknowledgments:} We are grateful to Sumit Das, Pei-Ming Ho, 
Samir Mathur and Jacek Pawelczyk for  discussions. This work was supported by 
the Department of Energy under contract DE-FG02-91ER40688-Task A.

 \bigskip
 \bigskip
\newsec{Appendix 1}
In this appendix we give the formulas and some derivations used in section 3.
For the computations for harmonics we use the following expressions:
\eqn\harmss{\eqalign
{ 
& {1\over 4\pi}\int_{S^2}{{x^{i_1}\dots x^{i_l}}\over \rho^l}=
{{ \partial_{J_{i_1}} \dots \partial_{J_{i_l}}\int d^3x e^{-{1\over2}x^2+Jx}|
_{J=0}}\over {4\pi \int_0^\infty d\rho \rho^{l+2} e^{-{1\over2}\rho^2}}}=\cr
& \qquad = {{\pi^{1\over2}}\over{2^{{l+2}\over2} \Gamma({{l+3}\over2})}}
(\delta^{i_1 i_2}\dots+\dots),\cr
}}
where $\dots$ mean all possible contractions between $i$'s. Then we obtain
$\prodtw$ from:
\eqn\twp{
\langle Y^I Y^J \rangle=\Omega^I_{i_1\dots i_{l_1}}
\Omega^J_{j_1\dots j_{l_1}}
{1\over 4\pi}\int_S^2{{x^{i_1}\dots x^{i_l} x^{j_1}\dots x^{j_{l_2}}}\over 
\rho^{l_1+l_2}}
}
and in a similar fashion $\prodthr$. The extra combinatorial factors
as $l!$ for $\prodtw$ and $\alpha_1! \alpha_2! \alpha_3!$ and $\Sigma!$
come from different combinatorial way to match the indices for $\Omega$'s.
\newsec{Appendix 2}
In this appendix, we derive the formulas used in section 4. In order to derive
the expression $\jcj$ we use the spin $1\over 2$ representation for $SO(3)$ 
rotations $e^{iJ^{1,2,3}x}$, replacing $x\rightarrow{\sigma\over2}$
with $\sigma$ being the Pauli matrices, and we obtain:
\eqn\rots{
cos({J\over2})+i{sin({J\over2})\over J}J\sigma=\prod_{k=1}^3
(cos({J^k\over2})+i{sin({J^k\over2})\over J^k}J^k\sigma)
} 
>From this, after straightforward algebraic manipulations we obtain $\jcj$.

For $\spre$, we define:
\eqn\dreftr{
K_{N+1}^l\equiv ({d \over d(cos J)})^l({sin(J(N+1))\over sin J})|_{J=0}    
}
and using $sin(J(N+1))=cos(JN) sinJ+sin(JN) cosJ$ we obtain the following
recurrence relations and conditions:
\eqn\dreftrr{\eqalign
{& K_{N+1}^l=K_N^l+(N+l) K_N^{l-1},\quad l \geq 1, \cr
 & K_N^0=N, \cr
 & K_1^l=0, \quad l \geq 1, \cr
}}

It is now easy to see that the expression in the RHS of $\spre$ satisfies 
$\dreftrr$.

\listrefs

\end